\documentclass {aa}
\usepackage{graphicx}
\usepackage{txfonts}
\begin{document}
   \title{Optical follow-up of BL Lac candidates in the 2BL sample}            

   \author{R. Nesci, S. Sclavi 
          \and
          E. Massaro
          }

   \offprints{R. Nesci}

   \institute{Dipartimento di Fisica, Universit\'a La Sapienza,
              P.le Aldo Moro 2, I-00185 Roma\\
              \email{roberto.nesci@uniroma1.it}          }

   \date{Received November 2004; accepted ...... }

\abstract{
We investigate the nature of the BL Lac candidates in the northern part
of the 2BL sample (\cite{Londish02}) searching for optical variability
by means of $R_C$ band photometry with the Asiago 1.8 and Loiano 1.5 meter
telescopes during years 2002, 2003 and 2004.
We also made aperture photometry of the candidates on the plates available 
on the web from the Palomar and UK Schmidt telescopes.
No significant variability was detected for the majority of the objects
without detected radio and/or X-ray emissions,
while clear variability was found for the few sources of the sample with
strong radio and/or X-ray emission, which is a constant feature of the 
"classical" BL Lacs.
Some considerations on the nature of the 2BL sources are made, taking into
account their most recent data available in proper motions catalogues 
(SuperCOSMOS, USNO-B1) and in the Sloan Digital Sky Survey.

}
\maketitle
%
%________________________________________________________________

\section{Introduction}
The history of the BL Lacertae objects started in 1968 (\cite{Schmitt}) with 
the discovery of the radio emission from the prototype of the class, 
the variable ``star'' BL Lac.
A few years later, after that some other similar objects have been found
(OJ 287, W Com, ON 325 and AP Lib), \cite{Strittmatter} made the first 
definition of the class.
Subsequent observations with X-ray satellites showed that BL Lac objects
were also remarkable X-ray sources.
The first attempts to derive a sample of BL Lacs starting from 
catalogues of optically variabile sources were unfruitful, while those based 
on Flat Spectrum Radio Sources provided soon a good deal of objects, 
indicating that the radio emission was a better discriminant. 
Several samples of BL Lac objects have been collected up to now, using 
different selection criteria (e.g. 1 Jy sample, \cite{Stickel};
Extended Medium Sensitivity Survey, \cite{Morris}, \cite{Wolter}; 
Rosat-Greenbank survey, \cite{Laurent-Muehleisen}; 
Sedentary Survey, \cite{Giommi}, DXRBS, \cite{Landt01}; 
HRX-BL Lac survey, \cite{Beckmann}).
In all these samples, the presence of a non thermal radio or/and X-ray emission
was required.

Nowadays, BL Lac objects are considered a subsample of the population of 
the Active Galactic Nuclei, and are generally defined as sources with 
featureless optical spectra, strong and fast variability in the radio, optical
and X-ray bands, strong and variable radio and X-ray emission and often with a 
high degree of polarization.
The precise definition of the BL Lac class is still a matter of debate: the
optical spectrum was defined as ``featureless'' if the equivalent widths of the
emission lines were less than 5 \AA \  and/or the CaII break was less than 
0.25 (\cite{Stocke}). This definition was later relaxed by \cite{Marcha}
who included in the BL Lac class also objects with stronger features 
(e.g. CaII break $\le$ 0.40). A recent discussion of  these criteria was made 
by \cite{Landt04}, who pointed out that the relative intensity of the AGN and 
of the host galaxy is a key factor for the optical line equivalent width.

Recently \cite{Londish02} defined the first ``optically selected''
sample of BL Lac candidates, the 2BL sample, as a by-product of the 2QZ sample
of QSO (\cite{Croom}). 
Two strips of sky were surveyed by the 2QZ project: one centered on the 
celestial equator and one at declination $-$30. The selection of the QSO 
candidates was made on UK Schmidt photographic plates scanned with the APM, 
searching for point-like sources with photographic $(u'-b)$ and $(b-r)$ 
Quasar-like colours. Spectra for these candidates were then
obtained with the 2dF fiber spectrograph mounted on the 3.9~m AAT telescope. 
BL Lac candidates were selected from these spectra, looking for objects in the 
magnitude range (19.97$\ge b \ge$18.25) and appearing featureless. 
High proper motion sources were then rejected as probable DC white dwarfs.

The remarkable feature of this sample is that the large majority of the 
sources (47 out of 56) have no radio emission at the NVSS (\cite{Condon})
sensitivity level (2.5 mJy) and that only 5 sources have been detected on the 
ROSAT All Sky Survey (RASS; \cite{Voges99}, \cite{Voges00}), all of them being 
radio loud. Only three of these RASS sources were already reported in the 
literature as BL Lacs.
The lack of X-ray and Radio emission in the large majority of the 2BL
sources is puzzling, because it implies an emission model somehow different 
from that generally accepted for BL Lacs.

The Spectral Energy Distribution (SED) of BL Lacs is indeed well represented, 
in the Log($\nu$)-Log($\nu F_\nu$) plane, by a double-bell shape. The first 
bump may be well described by a log-parabolic law (\cite{Landau}; 
\cite{Massaro04a}) and can be explained by a synchrotron radiation
from relativistic electrons moving in a jet aligned along our line of sight;
the second one, from X to $\gamma$-ray energies, is presently less well 
constrained in shape by the observations and is generally attributed to 
Inverse Compton scattering of low-energy photons on the same relativistic 
electrons.
According to the peak frequency of the first bump, 
BL Lac objects are further subdivided into HBL (high-energy peaked) and 
LBL (low-energy peaked), following the scheme introduced by \cite{GiomPad}.

The faintness (actually upper limits) of the radio/optical and X-ray/optical 
ratios of the 2BL objects means that the low-energy bump of the SED is much 
narrower than that of the ``typical'' BL Lacs. 
There is however no doubt about the extragalactic nature of a number of 
2BL sources: some of them have indeed a definite redshift from faint emission 
or absorption lines. The most notable case is 2QZ J215454.3$-$305654 
(\cite{Londish04}), a radio- and X-quiet source which appeared
featureless in the 2dF spectrum but showed a redshift $z=0.494$ in a high 
S/N spectrum obtained with the ESO-VLT. 
A thermal origin for the bulk of the emission (e.g. from an accretion disk) 
rather than from a synchrotron process might be an explanation, as suggested
by \cite{Londish04}:
this raises however the question whether such sources can still be properly 
classified as BL Lacs. 

To further explore the characteristics of the 2BL candidates we performed 
a search for optical variability, which is one of the defining features of the
class. In the optical range the LBL sources show a large variability 
(3-4 magnitudes), while HBL sources show a moderate variability range
(about half magnitude, see e.g. \cite{HeiWag}; \cite{KurNik}; \cite{Marchili}).
It is, however, unlikely that the 2BL candidates are of the HBL type: actually
their plot in the radio-optical and optical-X spectral index plane
($\alpha_{RO} - \alpha_{OX}$, \cite{Londish02} fig. 13) shows that all the 
X-ray undetected sources have $\alpha_{OX} \ge 1.0$, while HBL sources have 
$\alpha_{OX} \le 0.9$ (\cite{Beckmann}). We expect therefore to detect a
variation of the order of at least several tenths of magnitude if the 2BL
candidates behave like ``normal'' LBLs.

In this paper we report the results of our optical observations, covering
a time interval  of three years (2002-2004), of the 2BL sources well 
observable from northern mid-latitudes (the equatorial 2BL sample).
Furthermore we retrieved from on-line archives (ESO, STScI) a number
of the Palomar and UK-ESO Schmidt survey plates and derived photographic red
magnitudes to explore the historical behaviour of these sources.
Finally, we considered the Sloan Digital Sky Survey Data Release 2 
(SDSS-DR2, http://www.sdss.org/dr2/) photometric and proper motion data,
recently availables for all the sources of the equatorial sample.

In October 2003
Londish revised the 2BL sample in her PhD thesis (\cite{Londish03})
using, for a number of sources, updated proper motions from the SuperCOSMOS 
sky survey, 5 band photometry from the Sloan Digital Sky Survey Early 
Data Release (\cite{Stoughton}), her own JHK and optical photometry. 
These data allowed to recognize a number of candidates in the former 
(\cite{Londish02}) list as probable white dwarfs (from proper motion, 
variability or black-body spectral energy distribution). 
We present anyway in this paper our observational results for all the 
sources monitored by us, even if some of them are most likely stellar. 

\section{Observations}

\subsection{CCD observations}

The 2BL sample (\cite{Londish02}) consists of 56 objects.
We found that two sources in the equatorial strip, 2QZ J105534.36$-$012617 and
2QZ J121834.8$-$011955, both detected in the NVSS, were already classified as 
BL Lac objects. The former was identified because detected in the RASS
with a flux of $3.3\times10^{-13} erg \, s^{-1} cm^{-2}$ (\cite{Appenzeller}) 
while the latter was known as the optical counterpart of a radio source in the 
PKS catalogue (\cite{Condon76}).

Our follow-up observations were made during the years 2002, 2003 and 2004 in 
the $R_C$ band with the Asiago (1.8 m) and Loiano (1.5 m) telescopes,  
respectively equipped with the AFOSC and BFOSC instruments, covering a field 
of view slightly larger than 8 arcmin. Some Asiago observations were made in 
service mode. Weather conditions did not permit 
a uniform time sampling for all the objects. 
Table \ref{log} reports the dates of the observing runs.

    \begin{table}
      \caption[]{Observation log with the Asiago and Loiano telescopes.}
         \label{log}
$$
         \begin{tabular}{ccc}
            \hline
            \noalign{\smallskip}
Year &  Month, Day    & Observatory \\ 
\hline
2002 & February 11,18 & Asiago \\
2002 & March 7-10     & Loiano \\
2002 & March 19       & Asiago \\
2002 & May 13         & Asiago \\
2003 & February 24    & Asiago \\
2003 & March 22-25    & Loiano \\
2003 & April 23       & Asiago \\
2004 & February 14-15 & Asiago \\
2004 & February 20-24 & Loiano \\
2004 & March 23-24    & Loiano \\
2004 & March 29-30    & Asiago \\
2004 & April 26-29    & Asiago \\
\noalign{\smallskip}
\hline
\end{tabular}
$$
\end{table} 

Twentynine out the 31 sources in the equatorial strip were observed.
The typical integration time for the esposures was 10 minutes; two exposures 
were made for the fainter sources and then averaged. The three ``radio 
detected'' candidates of the equatorial strip (i.e. 2QZ J105534.36$-$012617, 
2QZ J121834.88$-$011956 and 2QZ J142526.20$-$011826) were initially not 
included in the monitoring list so our sampling for them is poorer. 
Two sources, 2QZ J114221.4$-$014812 (redshift $z$=1.276) and 
2QZ J114327.3$-$005050 (redshift $z=$1.591; \cite{Londish02}), 
were never observed.

Differential photometry of the BL Lac candidates was performed using a set of 
about ten stars in the field of each source, spanning a range of two or three 
magnitudes around the object of interest. These were selected so that 5 stars 
were nearly as faint as the target, to estimate the actual photometric accuracy
achievable (check stars), while the brighter ones were used to put on a common 
scale the instrumental magnitudes from different runs (reference stars). 
The zero point of the magnitudes for each field has been established adopting 
the GSC2 magnitude for the brightest reference star. A table of the adopted
reference stars and magnitudes is given in Appendix (Table A1).
For each frame a photometric radius equal to the average FWHM of the reference
stars was used. Aperture photometry was carried out using the DAOPHOT routine 
of the IRAF package. All the 2BL sources showed a point-like PSF in our images.

\begin{figure}
   \centering
   \includegraphics[width=9cm]{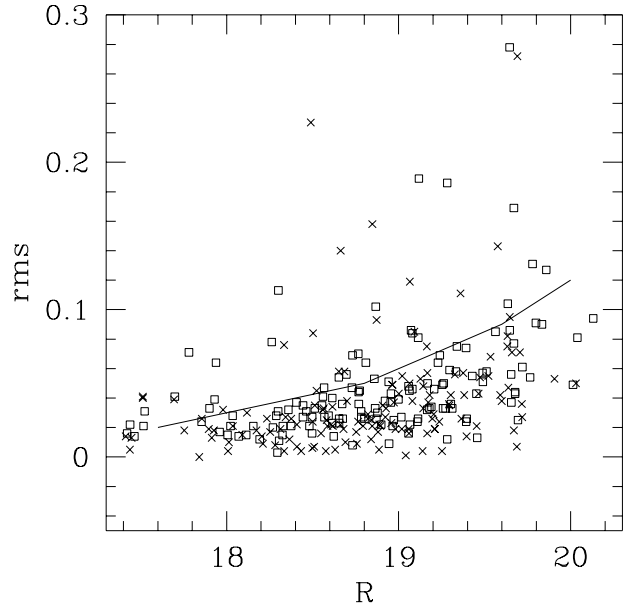}
   \caption{The rms deviation of the faint reference stars as a function of
   instrumental magnitude for the Loiano (squares) and Asiago (crosses) data.
   The continuous line marks the adopted boundary between true variability and
   photometric uncertainty (see text).}
    \label{rif}
\end{figure}

During the consistency check between the instrumental magnitudes obtained with 
the two telescopes, an appreciable systematic difference was found for a few 
stars: inspection of their colours using the public SDSS magnitudes showed 
that this difference was present only for ``red'' star (SDSS $r-i \ge$ 0.7). 
These stars were then discarded for the magnitude zero point definition in 
the final data reduction. 
In no field did we find more than one ``red'' star.
The typical average accuracy of our magnitudes as a function of a star 
luminosity can be derived from Fig.~\ref{rif}, where the rms deviation of the 
instrumental magnitudes for each check star of all our fields is plotted 
against its average magnitude, separately for the Asiago (crosses) and Loiano 
(squares) data sets. 
The general trend of this plot is of a larger dispersion for fainter stars, 
as expected.
Some stars however show a very large dispersion, so we considered them as 
variables and discarded them as check stars in the data analysis. 

To discriminate between true variability of a BL Lac candidate and 
photometric uncertainty we binned the magnitude values of the check stars
at steps of 0.4 mag and computed for each bin the average rms value and the 
dispersion of the data around this average. Then we assumed as
upper limit of the "statistical" deviations for each bin 
its average rms plus the dispersion. A smooth line approximating these
upper limits is shown in Fig.~\ref{rif} and was adopted as the border-line 
for detecting variability.

%%______________________________________________use the width command

Table~\ref{risul} gives the source name, the photographic red magnitude
derived from the $b$ and ($b-r$) values published by \cite {Londish02},
the average $R_C$ magnitude, rms deviation and number of 
observations for the Loiano and Asiago datasets and the border-line value of 
the photometric uncertainty.

    \begin{table*}
      \caption[]{Observed mean $R_C$ magnitudes and rms deviations for our Loiano and Asiago observations. The last column report the fiducial border line value, discussed in the text, for each source.}
         \label{risul}
$$	 
         \begin{tabular}{crccccccccccccc}	 
            \hline
            \noalign{\smallskip}
  Source name &2QZ $r$ &Loiano&rms&N& Asiago & rms&N &border-line\\ 
\hline
2QZ J100253.2$-$001728&19.03 &19.03&0.04&4&19.05&0.03&8&0.05\\
2QZ J102615.3$-$000630&19.17 &19.04&0.02&6&19.03&0.03&5&0.05\\
2QZ J103607.4$+$015658&18.57 &18.51&0.03&5&18.51&0.03&5&0.04\\
2QZ J104519.7$+$002615&18.43 &18.43&0.02&5&18.50&0.02&4&0.04\\
2QZ J105355.1$-$005538&19.09 &19.23&0.07&7&19.29&0.04&5&0.06\\
2QZ J105534.3$-$012617&18.50 &18.09&0.01&2&18.12&0.05&4&0.03\\ 
2QZ J110644.5$+$000717&$>$20.93 &20.33&0.25&5&20.39&0.05&6&0.14\\
2QZ J113413.4$+$001041&18.25 &18.18&0.02&7&18.26&0.05&4&0.03\\
2QZ J113900.5$-$020140&18.76 &18.72&0.06&5&18.74&0.03&4&0.04\\
2QZ J114010.5$-$002936&19.97 &19.88&0.09&9&19.96&0.07&6&0.12\\
2QZ J114137.1$-$002730&18.71 &18.85&0.06&7&18.95& -  &1&0.05\\ 
2QZ J114521.6$-$024758&18.53 &18.46&0.01&5&18.50&0.01&3&0.04\\
2QZ J114554.8$+$001023&19.46 &19.45&0.03&6&19.48&0.04&2&0.08\\
2QZ J115909.6$-$024534&18.77 &18.68&0.04&5&18.74&0.05&3&0.04\\
2QZ J120015.3$+$000552&18.63 &18.58&0.02&5&18.64&0.02&2&0.04\\
2QZ J120558.1$-$004216&18.46 &18.62&0.02&5&18.63&0.01&3&0.04\\
2QZ J120801.8$-$004219&18.86 &18.90&0.02&4&18.96&0.01&2&0.05\\
2QZ J121834.8$-$011955&18.25 &16.60&0.16&5&16.25&0.05&2&0.01\\   
2QZ J122338.0$-$015619&18.85 &18.96&0.01&5&19.04&0.03&1&0.05\\
2QZ J123437.6$-$012953&18.97 &18.73&0.05&5&18.76&0.02&1&0.04\\
2QZ J125435.7$-$011822&18.81 &18.87&0.04&4&     &    & &0.04\\  
2QZ J130009.9$-$022601&18.67 &18.77&0.03&6&18.78&0.01&3&0.04\\
2QZ J131635.1$-$002810&19.53 &19.44&0.03&6&     &    & &0.08\\
2QZ J132811.5$+$000227&19.79 &19.71&0.09&5&19.77&0.05&3&0.10\\
2QZ J140021.0$+$001955&19.30 &19.53&0.07&8&19.51&0.03&3&0.08\\
2QZ J140207.7$-$013033&20.14 &19.73&0.08&8&19.87&0.05&4&0.10\\
2QZ J140916.3$-$000012&18.47 &18.41&0.02&8&18.45&0.01&2&0.04\\
2QZ J141040.2$-$023020&19.27 &18.68&0.01&6&18.73&0.03&6&0.04\\
2QZ J142526.2$-$011826&19.47 &19.24&0.01&2&19.31&0.01&2&0.08\\  
\noalign{\smallskip}
\hline
\end{tabular}
$$
\end{table*} 

A first consideration from the data of Table~\ref{risul} is that, for the large majority of 
the sources, the average magnitudes in the Asiago and Loiano 
datasets are very similar, but the Loiano values are systematically 
slightly brighter, on average by 0.04 mag. Given that the Loiano and Asiago 
observations were made each year typically at an interval 
of about one month
(see Table~\ref{log}, it is unlikely that we observed our sources always in 
a brighter state at one telescope, so we attribute this difference mainly to 
the presence of a colour equation in the instrumental magnitudes rather than 
to intrinsic source variability. We applied therefore a systematic
correction of $-$0.04 mag to the Asiago (the less numerous) data quoted in 
Table~\ref{risul} before merging the two datasets.

A second consideration is that generally the rms deviations of both datasets
are smaller than the borderline derived from the comparison stars, highly 
suggestive of a limited, if any, optical variability, at variance with the 
expectations for BL Lac objects.

\begin{figure*}
   \centering
   \includegraphics[width=\textwidth]{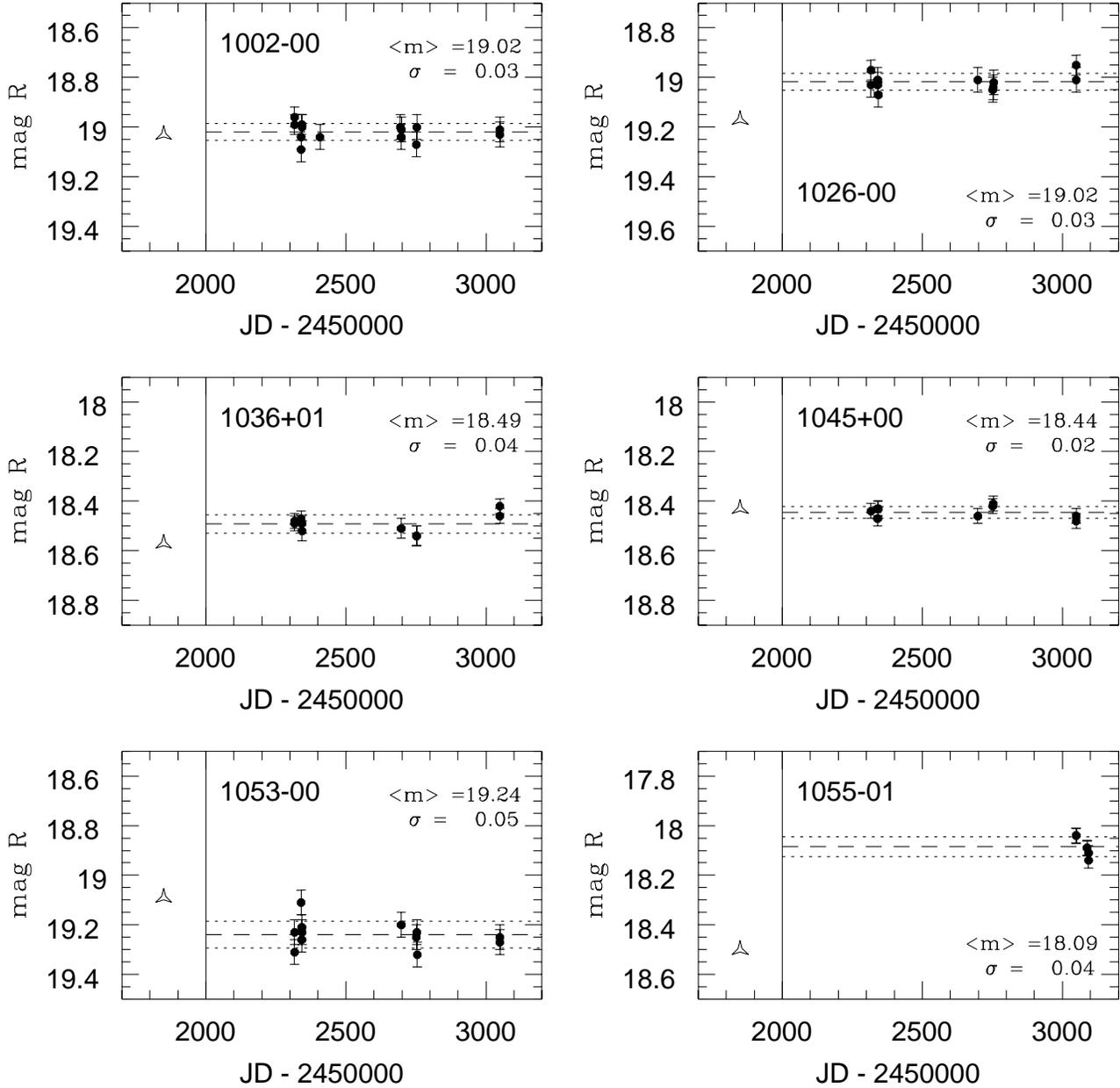}
    \caption{Light curves of 2BL candidates: abscissa is in Julian Date
$-$2450000, ordinate is $R$ mag. The triangle on the left of the vertical bar
indicates the value from \cite{Londish02} located at arbitrary JD. Black dots 
are our CCD data with error bars derived from the internal consistency of the
faint reference stars in the data set. The dashed line is the average value 
of the source in the data set; the dotted lines indicate the 
$\pm 1\sigma$ strip (see text).}
    \label{Fig1}
    \end{figure*}

\begin{figure*}
   \centering
   \includegraphics[width=\textwidth]{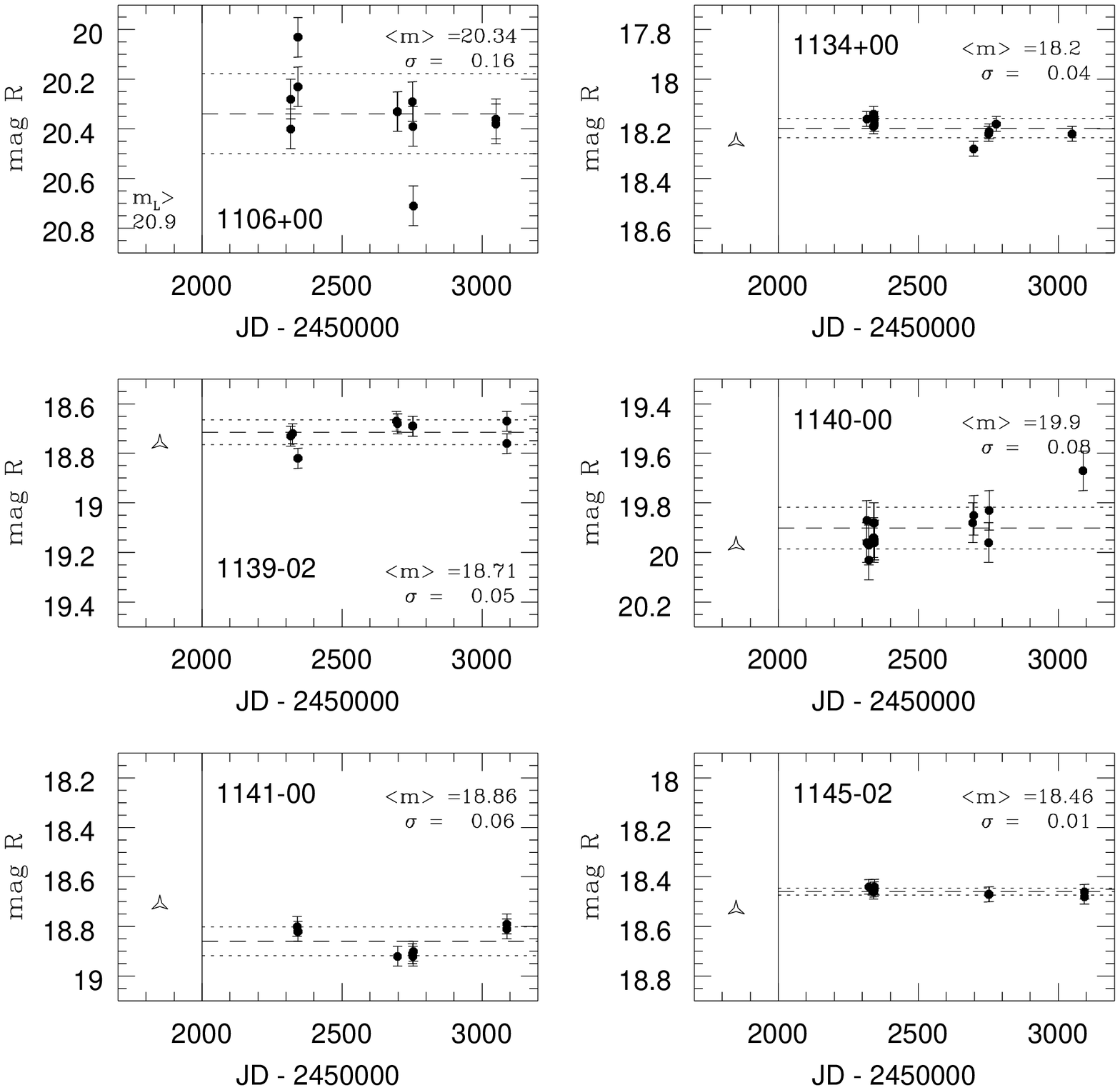}
    \caption{Light curves of 2BL candidates: abscissa is in Julian Date
$-$2450000, ordinate is $R$ mag. Symbols as in Fig. 2.}
    \label{Fig2}
    \end{figure*}

\begin{figure*}
   \centering
   \includegraphics[width=\textwidth]{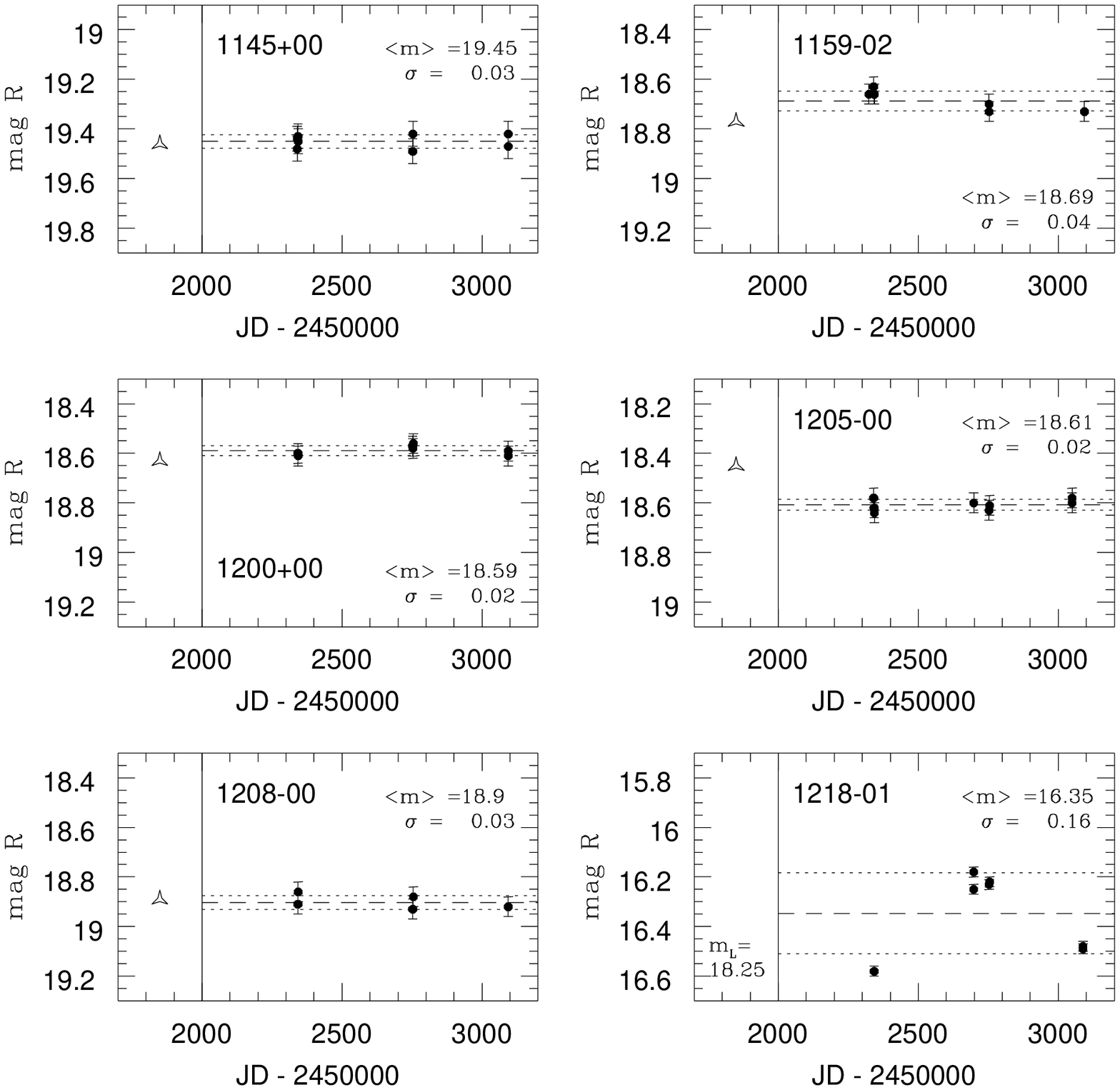}
    \caption{Light curves of 2BL candidates: abscissa is in Julian Date
$-$2450000, ordinate is $R$ mag. Symbols as in Fig. 2.}
    \label{Fig3}
    \end{figure*}

\begin{figure*}
   \centering
   \includegraphics[width=\textwidth]{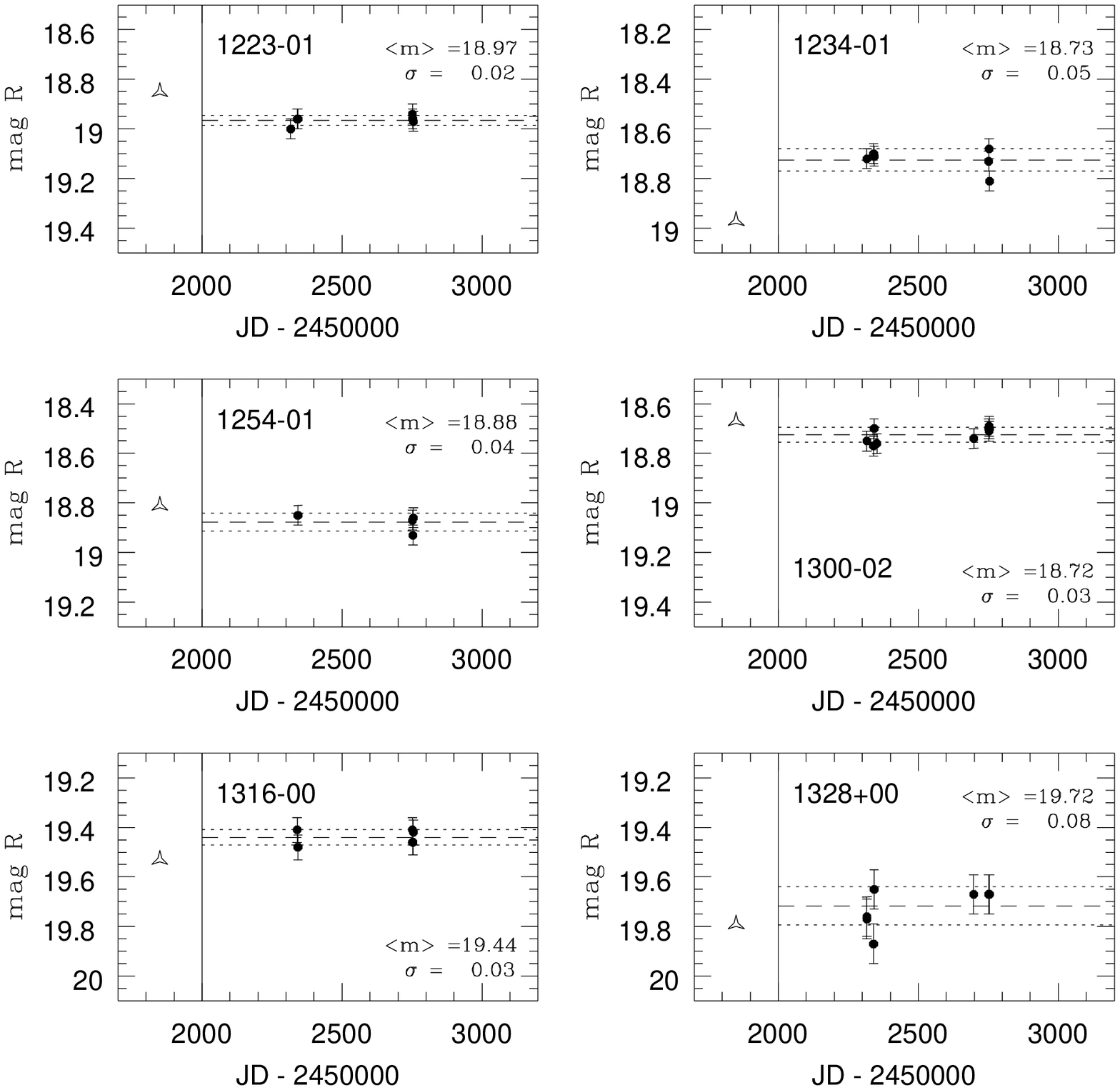}
    \caption{Light curves of 2BL candidates: abscissa is in Julian Date
$-$2450000, ordinate is $R$ mag. Symbols as in Fig. 2.}
    \label{Fig4}
    \end{figure*}

\begin{figure*}
   \centering
   \includegraphics[width=\textwidth]{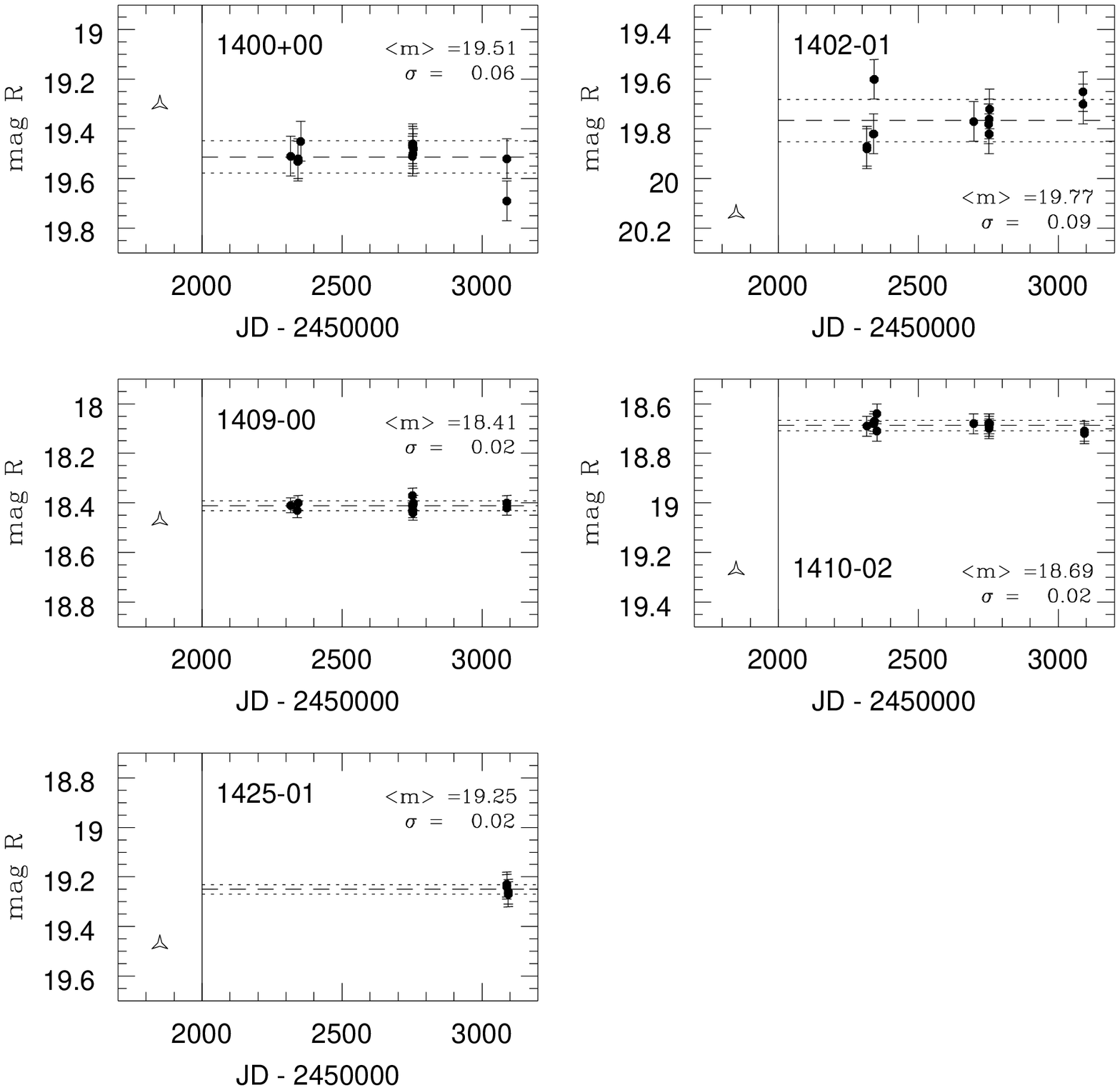}
    \caption{Light curves of 2BL candidates: abscissa is in Julian Date
$-$2450000, ordinate is $R$ mag. Symbols as in Fig. 2.}
   \label{Fig5}
   \end{figure*}

The resulting light curves are shown in Figs~\ref{Fig1} to~\ref{Fig5} where
we plot for each object our magnitudes and error estimates.
The average value is shown as a dashed line, while the range ($\pm 1 \sigma$)
of the distribution of the magnitudes of each source is shown by the dotted 
lines.

We found large variability only for two sources: 1218$-$01, a radio loud
(not X-ray detected) source, and 1106+00, which has a hot (32000 K) thermal
spectrum according to the SDSS photometry (\cite{Londish03}). 
The latter source is very faint ($ R \simeq$ 20.3, by far the faintest of our sample), 
and our photometry is rather inaccurate at this flux level (expected rms $=$ 0.14) so 
we are unsure of its variability.

A comparison of our data with the photographic 2QZ $r$ values in 
\cite{Londish02} (column 3 of our Table \ref{risul}) is not straightforward 
because of the possible presence of systematic effects in the magnitude scale 
for each field, and to the limited accuracy of photographic plates at these 
magnitudes. Both APM and GSC2 magnitudes have a formal error of 0.3 mag, which 
may be even larger at $r \simeq$ 19.5. We have checked for our ``bright'' 
($R \le$ 18.5) reference stars the consistency between GSC2 and our CCD 
magnitudes, finding always agreement within the formal accuracy of the GSC2.
Consistency between our magnitudes and those given in the on-line APM catalogue 
(http://www.ast.cam.ac.uk/$\sim$apmcat/) was good (i.e. within the formal 
error) for stars brighter than $R \sim$ 19, while strong differences were found 
for the fainter sources.
We decided therefore to consider as real only a difference greater than 0.3 mag
between our values and those derived from \cite{Londish02}.

Using the 2QZ $r$ photographic data, the variability of 1218$-$01 is confirmed 
and two more sources, 1055$-$01 (also with a NVSS radio detection) and 
1410$-$02 (radio quiet), are found to be varied. 
The variability of 1234$-$01, 1400+00, 1402$-$01 and 1425$-$01 is also 
possible, while we cannot get any firm conclusion regarding 1106+00 because
only a lower limit ($r \ge$ 20.9) can be derived from \cite{Londish02}.

%

%______________________________________________
%%______________________________________________ 
%%______________________________________________ 
%%______________________________________________ 

\bigskip

\subsection{Historical photographic data}
For each object we retrieved all the scanned images available on-line
coming from the POSS1 and POSS2 Surveys and the UK-Schmidt plates in the red 
(E or F emulsion) band.
Differential aperture photometry of the 2BL objects was made with the DAOPHOT 
routine of the IRAF package, using the same reference stars selected
for the CCD frames: the photometric radius was taken equal to the average FWHM
of the bright reference stars. 
Conversion from photographic magnitudes to our CCD $R_C$ magnitude scale was 
made building a calibration curve for each image by means of the reference 
stars: the CCD magnitude of the 2BL object was determined by interpolation 
between the nearest experimental points of this curve and it had an 
uncertainty of about 0.2 magnitudes. Only in a few plates 
was the object found to be apparently lower than our faintest reference star: 
in this cases we extended our CCD sequence to encompass the apparent magnitude of 
the candidate BL Lac on the plate.  

The historical (in our CCD-$R_C$ scale) magnitudes so derived are reported in
Table~\ref{Phot} for each object and epoch: the first column gives the source 
name shortened to the most significant digits of RA and DEC (HHMMsDD); 
subsequent columns give our estimated $R_C$ magnitude for each year. 
For some fields we found more than one plate in a given year, and their 
magnitudes are reported in subsequent lines. 
When the source was too faint on the plate to be reliably measured we report 
a lower limit.

    \begin{table*}
      \caption[]{Photographic magnitudes. Plates up to 1956 are from the POSS1 survey with 103aE emulsion, later plates are IIIaF from the POSS2 or UK Schmidt surveys.}
         \label{Phot}
         \begin{tabular}{lrrrrrrrrrrrrrr}
            \hline
            \noalign{\smallskip}
name   &1952&1955&1956&1985&1986&1987&1988&1989&1991&1994&1995&1996&1997&1998\\
\hline
1002-00&19.0&    &    &    &19.0&    &    &    &    &    &    &    &18.9&    \\
1026-00&19.6&    &    &    &    &19.3&    &    &19.1&    &    &    &    &    \\
1036+01&19.9&    &    &    &    &    &    &    &18.7&    &    &    &    &    \\
       &    &    &    &    &    &    &    &    &18.7&    &    &    &    &      \\
1045+00&    &18.2&    &    &    &    &    &    &18.6&    &    &    &    &    \\
1053-00&    &19.0&    &    &19.4&    &    &    &19.0&    &    &    &19.8&    \\
1055-01&    &18.8&    &    &18.5&    &    &    &    &    &    &    &17.6&    \\
1106+00&   &$\ge$20.0&    &    &20.1&    &    &    &    &    &    &    &20.5&    \\
1134+00&18.2&    &    &    &    &    &    &18.1&    &    &    &    &    &18.4\\
       &    &    &    &    &    &    &    &18.1&    &    &    &    &    &    \\
1139-02&19.1&    &    &    &    &    &    &    &    &    &    &    &    &19.2\\
1140-00&$\ge$19.8&    &    &    &    &    &    &    &    &    &    &    &   &$\ge$19.8\\
1141-00&$\ge$19.5&    &    &    &    &    &    &    &    &    &    &    &    &18.7\\
1145-02&18.5&    &    &    &    &    &    &    &    &    &    &    &    &18.5\\
1145+00&19.1&    &    &    &    &    &    &    &    &    &    &    &    &19.2\\
1159-02&    &19.0&    &    &    &    &18.8&    &    &    &    &18.8&    &    \\
       &    &    &    &    &    &    &    &    &    &    &    &18.8&    &    \\
1200+00&    &19.0&    &    &    &    &18.8&    &    &    &    &18.8&    &    \\
1205-00&    &18.5&    &    &    &    &18.5&    &    &    &    &18.5&    &    \\
1208-00&    &18.5&    &    &    &    &18.8&    &    &    &    &18.8&    &    \\
1218-01&    &17.6&    &    &    &    &    &18.6&    &    &    &    &18.5&    \\
       &    &18.1&    &    &    &    &    &    &    &    &    &    &    &    \\
1223-01&    &18.6&    &    &    &    &    &19.2&    &    &    &    &19.2&    \\
1234-01&    &18.5&    &    &    &    &    &18.6&18.9&    &    &    &18.6&    \\
       &    &    &    &    &    &    &    &    &19.1&    &    &    &    &    \\
1254-01&    &    &18.9&    &    &    &    &18.9&19.0&    &    &    &    &    \\
       &    &    &    &    &    &    &    &    &19.0&    &    &    &    &    \\
1300-02&    &    &19.2&18.8&    &    &    &18.9&    &    &    &    &    &    \\
1316-00&    & &$\ge$19.8&    &    &    &    &    &    &    &    &19.9&20.1&  \\
1328+00&    &    &19.5&    &    &    &    &    &    &    &    &19.6&19.5&    \\
1400+00&    &19.9&    &    &    &    &    &    &    &19.5&    &19.5&    &    \\
1402-01&   &$\ge$19.8&    &    &    &    &    &    &    &20.0&    &20.0&    &    \\
1409-00&    &18.7&    &    &    &    &    &    &    &18.6&    &18.5&    &    \\
1410-02&    &18.5&    &    &    &    &    &    &    &18.8&18.9&18.7&    &    \\
       &    &    &    &    &    &    &    &    &    &19.1&    &18.6&    &    \\
       &    &    &    &    &    &    &    &    &    &19.0&    &    &    &    \\
1425-01&    &20.0&    &    &    &    &    &    &    &19.7&    &19.4&    &    \\ 
\noalign{\smallskip}
\hline
\end{tabular}
\end{table*} 

Given the low accuracy achievable for the magnitudes of our sources from the 
Survey photographic plates, we assume that variability is believable only if 
a variation of at least 0.6 mag is detected. Small amplitude variability of 
our objects is therefore not testable in this way.
Only eight sources seem to be variable: 1036+01, 1053$-$00, 1055$-$01, 
1141$-$00, 1218$-$01, 1223$-$01, 1410$-$02 and 1425$-$01.
All the three radio loud sources are variable and two of them (1055$-$01 and 
1218$-$01) varied most significantly, as expected for classical BL Lac objects. 

\subsection{SDSS and proper motion data}

At the time of writing this paper (October 2004) all the objects in the 
equatorial strip of the 2BL sample (save the radio loud source 1055$-$01) have 
been observed photometrically (and several also spectroscopically) by the 
Sloan Digital Sky Survey (SDSS) and publicly available through the Data 
Release 2 (DR2).
%  (http://cas.sdss.org/astro/en/tools/chart/navi.asp).
This release covers a larger sky area and supersedes the Early Data Release 
(EDR, \cite{Stoughton}) used by \cite{Londish03}. 
The DR2 photometric observations include 5 bands ($u', g', r', i', z'$) 
covering the optical spectral range from 3500 to 9100 \AA, so that they 
allow us to check whether the spectral energy distribution is consistent with 
a black body (as expected for a DC white dwarf) or with a power law 
($F_{\nu}=A~\nu^{\alpha}$) which is typical of a BL Lac source. 
Typical values for $\alpha$ range between $-$1 and $-$2 for LBL sources (e.g. 
\cite{Vagnetti}), but may be flatter for HBL objects.

We performed this test for our sources, applying a correction of $-$0.04 to the 
published $u'$ magnitudes and +0.02 to the $z'$ magnitudes, as indicated in the
relevant SDSS web page. 
A correction for the galactic extinction was also made, following 
\cite{Schlegel}, which anyway is always small ($E_{(B-V)} \le$ 0.055), given 
the high galactic latitudes of the sources. To compute the $\chi^2$ of the fit 
we assumed an error of 0.03 mag for each photometric band, including any 
systematic effect.
The results came out to be largely independent of the assumed reddening 
correction and of the systematic corrections stated above: in no case
did a ``good''  black-body fit switch into
a power-law one (or vice/versa) including/excluding any of the above 
corrections. There are anyway a few cases with no good fit with either 
spectral shape, which remained ``bad'' also tuning the above corrections.
The values of spectral indices range between 0.0 and $-$1.4, which are 
reasonable for AGNs.
Our fits of the spectral shapes with a black body distribution give temperature 
values in fair agreement with the previous results by \cite{Londish03},
save 1106+00 which is very hot ($\sim$ 130,000 K). 

We report in Table \ref{sdss} several quantities for each source:
source shortened name (column 1), the SDSS $r$ magnitude (column 2),
the SDSS spectral classification (Sp. Cl., column 3 : Unclass means that 
the spectrum is too poor according to the SDSS team for a classification, 
while n.a. indicates a not available spectrum). 
In column 4 and 5 we report our best fit respectively for the spectral shape 
given as a temperature (in K) in case of a black body (BB) or as a 
spectral index $\alpha$ for a power law (PL).
Column 6 gives the $\chi^2$ of the fit: there are 3 degrees of freedom in the 
fit, so $\chi^2$ values greater than 9 means that the fit is acceptable only 
at a confidence level less than 2.5\%.
Columns 7, 8, 9  give the source proper motion according, respectively, to the 
SDSS, the USNO-B1 (\cite{Monet}) catalogue and the SuperCOSMOS Science Archive
(SSA, http://surveys.roe.ac.uk/ssa/): these motions are indicated in units of 
their rms deviation for the SDSS and SSA while for the USNO-B1 the proper 
motion probability is reported as given by the CDS, with 9 meaning more than 
90\%. 

\begin{table*}
\caption[]{Photometric, spectral parameters and proper motion significance for the monitored  sources. }
\label{sdss}
$$
\begin{tabular}{cccrrrrcr}
\hline
\noalign{\smallskip}
 Source    & $r$ & Sp. Cl. & BB Temp. & PL $\alpha$ &$\chi^2$ &  \multicolumn{3}{c}{proper motion significance} \\
           & mag &         & K        &              &         & SDSS & USNO-B1 & SSA \\
\hline
 1002$-$00 & 19.17&n.a.        & 10800  &        &    4.0 &  9.81& 9& 12.47\\
 1026$-$00 & 19.28&n.a.        & 11900  &        &    2.5 &  6.08& 9&  5.69\\
 1036$+$01 & 18.80&QSO z$=$1.86&        & $-$0.28&    3.1 &  0.50& 0&  0.12\\
 1045$+$00 & 18.62&STAR        & 11900  &        &    0.8 &  6.78& 9&  7.61\\
 1053$-$00 & 19.42&n.a.        & 10200  &        &   11.2 &  2.26& 8&  2.18\\
 1106$+$00 & 20.46&n.a.        &130000  &        &  238.3 &  0.61& 0&  0.27\\
 1134$+$00 & 18.43&QSO z$=$1.48&        & $-$0.71&   11.9 &  0.41& 0&  0.25\\ 
 1139$-$02 & 19.18&GAL?        &        & $-$1.42&    5.4 &  1.20& 0&  0.40\\
 1140$-$00 & 20.02&STAR        & 13100  &        &   16.6 &  1.95& 0&  3.60\\
 1141$-$00 & 18.64&Unclass     &        & $-$0.61&   16.4 & 21.45& 9&  0.67\\
 1142$-$01 & 18.20&n.a.        &  7300  &        &   24.8 &  0.50& 0&  0.00\\
 1145$-$02 & 18.67&Unclass     & 11300  &        &    2.3 &  1.40& 9&  1.97\\
 1145$+$00 & 19.57&n.a.        & 11300  &        &    2.7 &  1.21& 0&  2.29\\
 1159$-$02 & 19.03&Unclass     &        & $-$0.39&   16.1 &  0.56& 9&  0.50\\
 1200$+$00 & 19.07&QSO z$=$1.65&  6300  &        &   91.0 &  0.46& 0&  0.56\\
 1205$-$00 & 18.74&n.a.        &  7300  &        &    7.1 & 11.71& 9& 13.23\\
 1208$-$00 & 19.08&n.a.        & 12400  &        &    4.5 &  2.26& 9&  2.58\\
 1218$-$01 & 17.56&Unclass     &        & $-$1.33&    0.1 &  0.68& 9&  0.58\\
 1223$-$01 & 19.12&n.a.        & 13100  &        &   20.5 &  2.34& 9&  2.55\\
 1234$-$01 & 19.17&n.a.        &  7600  &        &   30.8 &  0.68& 0&  1.45\\
 1254$-$01 & 19.16&n.a.        &  7600  &        &    6.3 &  1.88& 0&  2.42\\
 1300$-$02 & 18.93&Unclass     &  8000  &        &   11.5 &  9.09& 9& 10.48\\
 1316$-$00 & 19.60&n.a.        &        & $-$0.58&   26.2 &  3.11& 9&  9.06\\
 1328$+$00 & 19.77&n.a.        & 10800  &        &    4.0 &  2.44& 9&  2.52\\
 1400$+$00 & 19.65&n.a.        &        & $-$0.03&    4.4 &  0.19& 0&  3.86\\
 1402$-$01 & 19.88&STAR        & 15100  &        &   45.1 &  1.08& 9&  0.43\\
 1409$-$00 & 18.66&Unclass     &  8800  &        &    4.6 &  3.52& 8&  3.84\\
 1410$-$02 & 18.97&Unclass     & 12400  &        &   16.4 & 10.69& 9&  1.13\\
 1425$-$01 & 19.08&Unclass     &        & $-$0.98&    5.6 &  0.11& 0&  0.97\\
\noalign{\smallskip}
\hline
\end{tabular}
$$
\end{table*} 

An approximate conversion of the SDSS $r$ magnitudes into our GSC2-based $R_C$
magnitudes may be made assuming $R_C=r-$0.23 (see \cite{Fukugita}). 
Given that the zero points of our magnitude scale for our fields may have a 
systematic offset of about 0.2 mag, we considered as variable
only those sources which have a magnitude difference, with respect to our 
Loiano values, $\ge$ 0.3 mag.
Three objects appear to be variable from this comparison, 1141$-$00, 1218$-$01 
and 1425$-$01, the last two being radio loud; all these 3 objects appear 
variable also in the historical survey plates.

    \begin{table*}
      \caption[]{ Summary of variability detection and radio flux
	densities of the monitored sources. Sources with the flag X
	were rejected by Londish (2003) because of high proper motion.}
         \label{variab}
         \begin{tabular}{cccccccccc}
            \hline
            \noalign{\smallskip}
 & Source    & Our monitoring & Londish & On line plates & SDSS & \multicolumn{2}{c}{Radio data} &  Flag \\
 &           &                &         &                &      &  $F(1.4 GHz)^1$ & $F(8.4 GHz)^2$         &       \\ 
 &           &                &         &                &      &   mJy      &   mJy             &       \\
\hline
 & 1002$-$00 & no       & no       &  no      & no  &     &           & X \\
 & 1026$-$00 & no       & no       &  no      & no  &     &           & X \\
 & 1036$+$01 & no       & no       & yes      & no  &     & $\le$ 0.1 &   \\
 & 1045$+$00 & no       & no       &  no      & no  &     &           & X \\
 & 1053$-$00 & no       & no       & yes      & no  &     & $\le$ 0.1 &   \\
 & 1055$-$01 & no       & yes      & yes      &  -  & 11  &           &   \\ 
 & 1106$+$00 & possibly & ?        & no       & no  &     & $\le$ 0.1 &   \\
 & 1134$+$00 & no       & no       & no       & no  &     &           &   \\
 & 1139$-$02 & no       & no       & no       & no  &     &           &  \\
 & 1140$-$00 & no       & no       & no       & no  &     &           & X \\
 & 1141$-$00 & possibly & possibly & yes      & yes &     &   3.6     &   \\ 
 & 1145$-$02 & no       & no       & no       & no  &     &           & X \\
 & 1145$+$00 & no       & no       & no       & no  &     &           &   \\
 & 1159$-$02 & no       & no       & no       & no  &     &   0.3     &   \\
 & 1200$+$00 & no       & no       & no       & no  &     &           &   \\
 & 1205$-$00 & no       & no       & no       & no  &     &           & X \\
 & 1208$-$00 & no       & no       & no       & no  &     &           & X \\
 & 1218$-$01 & yes      & yes      & yes      & yes & 244 &           &   \\   
 & 1223$-$01 & no       & no       & yes      & no  &     &           & X \\
 & 1234$-$01 & no       & possibly & no       & no  &     &           &   \\
 & 1254$-$01 & no       & no       & no       & no  &     &           &   \\  
 & 1300$-$02 & no       & no       & no       & no  &     &           & X \\
 & 1316$-$00 & no       & no       & no       & no  &     &           & X \\
 & 1328$+$00 & no       & no       & no       & no  &     &           & X \\
 & 1400$+$00 & no       & no       & possibly & no  &     &  0.3      & X \\
 & 1402$-$01 & possibly & possibly & no       & no  &     & $\le$ 0.1 &   \\
 & 1409$-$00 & no       & no       & no       & no  &     &           & X \\
 & 1410$-$02 & no       & yes      & yes      & no  &     &           &   \\
 & 1425$-$01 & no       & possibly & yes      & yes &  10 &           &   \\  
\noalign{\smallskip}
\hline
\end{tabular}

1.~ Flux density from NVSS.

2.~ Flux density from Londish et al. (2002), Londish (2003) 
\end{table*} 

\section{Variability properties}

Flux variability at optical frequencies is one of the characteristics of the 
BL Lac objects.
The most monitored sources are generally LBL sources which show no definite 
time scale and a range of variability of a few magnitudes.
HBL sources are generally considered to be less variable in the optical; 
however most of the well studied HBL's like the so-called TeV blazars 
(e.g. Mkn 421, Mkn 501, 1ES 2344+514, 1ES 1959+650) are relatively nearby 
sources with a strong host galaxy, so that the apparent luminosity variation 
in the optical is reduced by the steady stellar component of the integrated 
flux. As told in the Introduction, however, our sources are not likely to 
be HBL, given the undetection of their X-ray flux.

We collect in Table~\ref{variab} our evaluations of the variability of each 
source (column 1) in our data (column 2), with respect to the photographic 
2QZ data (column 3), to the older historical data (column 4) and to the SDSS 
photometry (column 5). 
In columns 6, 7 we report the radio flux in mJy respectively at the 
frequencies of 1.4 and 8.4 GHz according to \cite{Londish02} and 
\cite{Londish03}.
The flag X in the last column marks objects excluded from the 2BL sample by
\cite{Londish03}.

The results of our variability search may be summarized as follows:

i) only one source (1218$-$01, radio-loud) showed large variability in our 
CCD monitoring (we will discuss on it in more detail at the end of the 
section);

ii) three sources (two of them radio-loud) showed variability when 
the 2QZ photographic data are compared with our CCD data;

iii) eight sources (including all the 3 radio-loud) showed variability when 
the public digitized Sky Survey plates (spanning a 50 years interval) are 
measured using our reference CCD sequences.

iv) three sources showed variability comparing our CCD data with those of the 
SDSS, two of them being radio-loud (the third radio-loud source is still not 
observed by the SDSS).

This gives a total of eight variable sources out of the 29 observed, as 
indicated in Table~\ref{variab}. Notably, 
none of the 3 sources with a spectroscopically confirmed stellar spectrum 
showed variability in our study.

Of these 8 sources:

a) three are radio loud (1055$-$01, 1218$-$01, 1425$-$01); 
the first two were also detected in the J, H and K bands by the 2MASS
survey, with magnitudes 16.7, 16.0, 15.3 (1055$-$01) and 15.2, 14.4,
13.5 (1218$-$01); the last two have 
also SDSS 5-band photometry which give unambiguous power law SEDs.

b) one is a radio-quiet QSO (1036+01) with a power law spectrum;

c) one has a power law spectrum (1141$-$00) and a faint radio detection, but 
   strong proper motion in the SDSS/USNO-B1;

d) three have a possible hot black body spectrum (1053$-$00, 1223$-$01 and 
   1410$-$02) and also a possible proper motion detection: our power-law fit 
   gives for all these sources $\chi^2$ values much worse then the black-body. 
   1223$-$01 was excluded from the sample by \cite{Londish03} due to its proper 
   motion; 1410$-$02 has a strong proper motion in the SDSS/USNO-B1 catalogues.

From our combined variability and spectral shape study, therefore, it seems 
that all the 3 radio loud sources among the 2BL candidates in the equatorial 
strip have a variability behaviour similar to ``classical'' BL Lac objects. 
We remark that only 4 of the sources found to be variable are 
definitely of extragalactic nature.

What about the 21 non-variable sources (at least at our sensitivity level)?
Twelve of them have already been excluded by the 2BL sample by 
\cite{Londish03}; our reanalysis of the spectral energy distribution and 
proper motions confirm that they are most likely stellar sources 
(see Table~\ref{variab}).
Of the remaining 9, one is a spectroscopically confirmed star (1402$-$01); 
2 are QSOs (1134+00 and 1200+00); 2 have a black-body spectral shape and 
probable detected proper motion (1145+00 and 1254$-$01); only two have a 
power-law spectrum (1139$-$02 and 1159$-$02), the second one with a faint 
radio detection at 8.4 GHz (\cite{Londish03}). 
Finally, 1234$-$01 has an intervening absorber at z$=$1.06, according to 
\cite{Londish03}, and is therefore most likely an extragalactic source.

Only 5 of the non-variable sources are surely or probably extragalactic. 
Summed to the 4 variable ones, we get a total of 9 AGNs, some of them with 
unclear nature.
We have checked that the equivalent width of the emission lines in the
sources classified as QSO in the SDSS are much larger ($\sim$ 40 \AA) than the 
classical 5 \AA \ limit by Stocke et al. (1991) or the extended limit of 
$\sim$30 \AA ~ by \cite{Marcha} and should not be considered 
therefore as BL Lac objects.

A remarkable case is that of 1218$-$01. It was identified by Condon et al.
(1976) as the optical counterpart of the radio source PKS B1216-010 with
a featureless spectrum (Wilkes et al. 1984). UBV photometry (Adam 1985)
indicates that it was even brighter than in our observations, and
considering the faint state reported by Londish et al. (2002) one can estimate
a variation range of at least 3 magnitudes.
A redshift $z=0.415$ is reported in the literature after Downes et al.
(1985) who quote a private communication never confirmed by any 
subsequent published line identification. Radio images from VLBA 
calibrator survey show a compact bright core (Beasley et al. 2002).
It is therefore a very typical bright BL Lac object included in the 2BL 
sample because of an occasional faint state.

\section{Conclusion}

The main problem we faced on in this work is how to identify and classify 
BL Lac objects. Of course, the mere
observation of a featureless optical spectrum is not sufficient 
although very specific of this class of AGNs. 
Detection of a significant variability can be a further request, however 
it could be difficult to be noticed because of the possible occurrence 
of quiescent periods and of the randomness of observation sampling.
Although the failure in detecting a variability does not necessarily 
imply that a source is not a BL Lac objects, the fact that we
observed significant brightness changes mainly in radio loud sources
indicates that the search for variations can be a useful strategy to 
confirm their nature.

It is an open problem if a population of ``radio quiet'',
 i.e. with a radio-optical spectral index less than 0.2 (Giommi 
\& Padovani 2004) BL Lac objects does really exist. 
At present, evidence for such a type of AGNs is elusive and the 
2BL sample must be considered partially contaminated by 
other types of objects, mainly of stellar nature.
We tried to establish what should be the SED of a source to be properly 
considered as a ``radio quiet'' BL Lac. It is known that the synchrotron 
emission in many LBL objects covers the frequency interval from radio to the 
optical-UV: the energy distribution is generally well represented by a 
log-parabola (\cite{Landau}, \cite{Massaro04a}), also combined with 
a power law on the low frequency side (\cite{Massaro04b}):
\begin{equation}
F(\nu) = A \nu^{-[a + b Log(1+ \nu/\nu_1)]}
\end{equation}

\begin{figure}
   \centering
   \includegraphics[width=8.5cm,height=8.5cm]{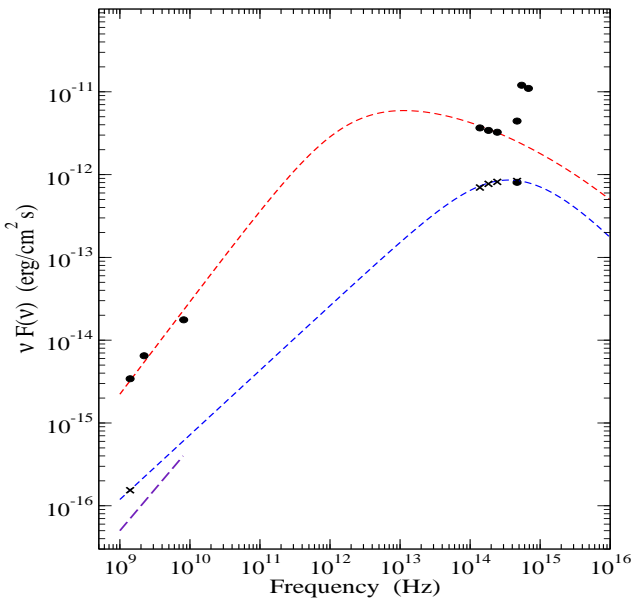}
   \caption{The Spectral Energy Distributions (SED) of 
             two radio loud BL Lac objects in the 2BL sample: 1218$-$01 
             (filled circles) and 1055$-$01 (crosses) derived from our photometry, 2MASS 
             and other literature data. Optical points of 1218$-$01 are also
             representative of its variability. Short dashed lines are the SEDs 
             modelled with eq.(1) and the long dashed line corresponds to a flat 
             radio spectrum at a flux density of 5 mJy.
           }
    \label{sed}
\end{figure}
%%________________________________________

Fig. \ref{sed} shows two examples of SEDs of two radio loud BL Lac objects in 
the 2BL sample. The peak frequency of the brighter (1218$-$01) is around  
10$^{13}$ Hz while that of 1055$-$01 is in the optical at about 4$\times$10$^{14}$ 
Hz. Of course they are based on not simultaneous data and therefore must be
considered only indicative of the actual distributions. 
Note that if the power-law branch of 1055$-$01 would be only moderately flatter 
than the one plotted in Fig. \ref{sed}, its radio flux density would be  
lower than 5 mJy, making the source only marginally detectable in many
radio surveys.
According to this picture a characteristic of most 2BL sources is that their
synchrotron peak must be at frequencies higher than 10$^{14}$ Hz, which
correspond to a flat optical spectrum. Indeed we see from Table \ref{sdss}
that all the 
sources with a PL spectrum, with the only exception of 1139$-$02, have 
a spectrum flatter than that of 1218$-$01 and of typical BL Lacs, with best
fit spectral indices greater than $-$1.0. 
Consequently, the frequency of their synchrotron peak must be in the 
optical-UV range. 
We expect therefore that these sources, differently from 1055$-$01 and 
1218$-$01, would not be very bright in the near IR: in particular,
$R_C-K$ colour indices could be smaller than typical BL Lac 
values (i.e. between 3 and 4, e.g. \cite{Nesci03}), making them too faint 
to be detectable in the 2MASS survey, which has a K-band limit of 14.5 mag 
(\cite{Skrutskie}). 
In any case, near IR follow-up of these sources would be very helpful to 
clarify their nature. 

The detection of sources with featureless optical spectrum, without a
radio and X-ray counterpart, can be a tool to select candidate BL Lacs, 
however its efficiency cannot be high because of
contamination from other galactic and extragalactic objects.
A population of ``optically'' selected BL Lac objects, with 
characteristic similar to the 2BL sample, should have peculiar
spectral properties implying different emission, geometrical and
physical conditions with respect to normal BL Lacs.  
A small number of sources could actually show these characteristics, and 
further multiwavelenght follow-up of this sample is required to unambiguously 
pick up them.  Our variability search, in fact, indicated that a
fraction of 2BL sources are not genuine AGNs. 
We expect that, in addition to variability studies like that presented
in this paper, polarimetric measurements can be another useful
selection tool as already successfully applied to radio
sources by Impey \& Tapia (1988), who discovered new optical faint
blazars at high redshift.

\begin{acknowledgements}
We thank Silvano Desidera and Hripsime Navasaradyan for the service-mode 
observations at the Asiago Observatory.
We are also indebted with Corinne Rossi for performing the observations 
in the February 2004 Asiago run.
We grateful to Paolo Giommi and Dario Trevese for useful discussions.
This paper makes use of the Sloan Digital Sky Survey Data.
Part of this work was supported by the italian MIUR under grant Cofin 
2001/028773 and 2003/027534.
\end{acknowledgements}

APPENDIX A

In this Table we give, for each monitored source, the adopted R magnitude of the reference star, used to bring onto a common scale the instrumental magnitudes obtained in the different runs from the Asiago and Loiano telescopes.

Column 1 is the shortened source name, column 2 the star identification in the GSC2 catalogue, column 3 the Right Ascension, column 4 the Declination and column 5 the adopted magnitude as reported in the GSC2.

\begin{table}
\caption[]{Standard stars used for the magnitude scale definition of each observed field}
\label{std}
$$
\begin{tabular}{clccc}
\hline
\noalign{\smallskip}
 Field     &  GSC2 name   & RA(2000)     & DEC(2000)      & $R$ mag \\
\hline                       
 1002$-$00 & S12121203157 & 10 03 07.247 & $-$00 20 52.80 & 15.88   \\
 1026$-$00 & S12120004610 & 10 25 54.636 & $-$00 04 05.29 & 15.81   \\
 1036$+$01 & N20310011951 & 10 36 03.198 & $+$01 57 50.01 & 16.67   \\ 
 1045$+$00 & N20020123674 & 10 45 27.082 & $+$00 26 38.51 & 16.63   \\
 1053$-$00 & S12013102499 & 10 53 59.120 & $-$00 57 07.98 & 15.78   \\
 1055$-$01 & S12013134345 & 10 55 45.967 & $-$01 26 38.83 & 15.83   \\
 1106$+$00 & N2002102736  & 11 06 44.789 & $+$00 04 23.30 & 15.56   \\
 1134$+$00 & N2000210760  & 11 34 24.908 & $+$00 09 22.80 & 16.25   \\
 1139$-$02 & S12003215130 & 11 38 55.494 & $-$02 03 48.55 & 16.34   \\
 1140$-$00 & S12000102281 & 11 40 18.739 & $-$00 30 28.04 & 16.65   \\
 1141$-$00 & S12000102432 & 11 41 38.253 & $-$00 28 24.10 & 16.59   \\
 1145$-$02 & S1200322301  & 11 45 19.259 & $-$02 48 24.62 & 15.96   \\
 1145$+$00 & N20000211863 & 11 45 50.451 & $+$00 14 28.74 & 15.81   \\
 1159$-$02 & S1200031962  & 11 59 21.166 & $-$02 43 03.89 & 15.11   \\
 1200$+$00 & N12000001088 & 12 00 22.134 & $+$00 06 25.84 & 16.42   \\
 1205$-$00 & S20000021747 & 12 05 55.459 & $-$00 40 07.58 & 16.13   \\
 1208$-$00 & S20000021893 & 12 08 00.074 & $-$00 38 42.07 & 15.10   \\
 1218$-$01 & S20003101017 & 12 18 36.504 & $-$01 16 31.96 & 16.47   \\
 1223$-$01 & S20002112306 & 12 23 42.524 & $-$01 53 01.92 & 15.85   \\
 1234$-$01 & S20002021494 & 12 34 40.338 & $-$01 29 28.60 & 15.27   \\
 1254$-$01 & S20021311531 & 12 54 34.142 & $-$01 15 54.62 & 15.39   \\
 1300$-$02 & S20023213907 & 13 00 19.358 & $-$02 24 31.19 & 16.35   \\
 1316$-$00 & S20020122684 & 13 16 41.468 & $-$00 27 21.27 & 16.28   \\
 1328$+$00 & N1201000669  & 13 28 22.040 & $+$00 03 58.83 & 16.29   \\
 1400$+$00 & N12121213738 & 14 00 15.893 & $+$00 23 41.79 & 15.98   \\
 1402$-$01 & S2021232715  & 14 02 06.761 & $-$01 26 44.35 & 16.42   \\
 1409$-$00 & S20212017779 & 14 09 07.170 & $-$00 00 37.65 & 16.02   \\
 1410$-$02 & S20232014778 & 14 10 42.436 & $-$02 27 13.09 & 16.27   \\
 1425$-$01 & S20201311703 & 14 25 20.602 & $-$01 18 56.53 & 16.13   \\
\noalign{\smallskip}
\hline
\end{tabular}
$$
\end{table}

\end{document}